\documentclass{article}
\usepackage{spconf,amsmath,graphicx}
\usepackage{enumitem}
\usepackage{amsfonts}

\title{Transcoding a 3D Gaussian Splatting Model from a Plenoptic Point Cloud or Mesh without the Original Multi-view Images}

\name{Maja Krivoku\'{c}a$^{\star}$ \qquad Riad Bendouro$^{\dagger}$ \qquad Neus Sabater$^{\star}$}
\address{$^{\star}$InterDigital France \\
$^{\dagger}$Centre Inria de l'Université de Rennes}

\begin{document}
\ninept
\maketitle
\begin{abstract}
In this paper, we propose an end-to-end \textit{transcoding} pipeline, to create 3D Gaussian splatting (3DGS) models from existing 3D plenoptic point cloud or mesh models, when the original multi-view images of the captured 3D object or scene are not available. We also propose a custom initialisation to guide the 3DGS model learning, with constraints to ensure that the final 3DGS model aligns closely with the input point cloud or mesh surface. Tests on a high-quality, standard plenoptic point cloud dataset show that our pipeline produces 3DGS models of high visual quality, with many fewer splats than points in the original dense point clouds. Additionally, our custom initialisation leads to much faster convergence and cleaner surface representation than when starting from the default SfM-based initialisation that is typically used for 3DGS model learning.
\end{abstract}
\section{Introduction}
\label{sec:intro}

Over the past two years, \textit{3D Gaussian Splatting} (3DGS)~\cite{Kerbl2023} has rapidly (re-)gained popularity as a representation and rendering method for producing photo-realistic digital versions of captured real-world objects or scenes. A 3DGS model represents a scene as a collection of 3D ellipsoids, or \textit{splats}, each of which is parameterized as a Gaussian function that allows the splat to be stretched, scaled, and rotated by different amounts and in different directions in 3D space. Each splat also has colour values associated with different viewing directions in 3D space, as well as an opacity coefficient that controls the splat’s transparency. This makes 3DGS particularly useful for modelling scenes that contain objects with non-Lambertian surface properties (e.g., shiny, reflective surfaces), whose colours may appear different when viewed from different viewing directions in 3D space. During rendering to a 2D display, the value of each pixel in the output 2D image is obtained by blending, in depth order, different intersecting Gaussian splats, taking into account their opacity values to produce a realistic final output. Typical generation methods for 3DGS models rely on having access to a (large) number of captured images of the 3D object or scene of interest, then learning a 3DGS model from those images. However, in some applications, the captured content may already be available in a more traditional volumetric media format, such as a 3D point cloud or mesh, and the original multi-view images that were used to produce this 3D model may be unavailable; but we may wish to create a 3DGS representation of this data. For example, the MPEG \textit{plenoptic point cloud} dataset, the “8i Voxelized Surface Light Field” (8iVSLF)~\cite{Krivokuca2018}, is a high-quality, dense point cloud dataset obtained from real captures, containing view-dependent colours per point, but without the original 2D images being available. Given the popularity of point cloud (and mesh) representations in recent years (including in MPEG codec standardisation efforts), and now the emerging popularity of 3DGS representations, it is natural to wonder how a 3DGS version of a dataset such as the 8iVSLF would compare to its original, plenoptic point cloud version, and whether there would be advantages in representing such data in 3DGS form instead. In this paper, we therefore propose methods to generate a 3DGS model of a captured 3D object or scene when starting from an already-available plenoptic point cloud (or mesh) of this object or scene, and when the original multi-view images that were captured of the scene are not available. We have not found similar ideas presented in the existing 3DGS literature. In~\cite{Hu2024}, a system is proposed for rendering a 3D point cloud into 2D images from arbitrary views without relying on the captured input images, but using a convolutional neural network to learn the 3DGS model. This is different to our system, which uses no neural networks for learning. Some works (e.g.,~\cite{Waczynska2024,Guedon2024,Choi2024}) propose methods to align 3DGS models with mesh surfaces, but these all rely on either the 3DGS model or the captured multi-view images already being available as input, whereas we do not.

We use the 8iVSLF dataset~\cite{Krivokuca2018} to explain and evaluate our ideas in this paper. This is currently the only publicly available dataset of \textit{real volumetric captures} that also contain \textit{view-dependent} attributes. Since the 8iVSLF dataset is of a high quality and closely related to 3DGS data, we believe that transcoding such plenoptic data to a 3DGS format is valuable. Our ideas in this paper may also be applied to \textit{non-plenoptic} point clouds or meshes (i.e., those without view-dependent colours); however, this would be less interesting for the purposes of 3DGS work, since a key reason for using a 3DGS model is to represent objects whose appearance varies with the viewer's point of view.

\section{Overview of 3DGS Model Learning and Introduction to Plenoptic Point Clouds}
\label{sec:background}

\begin{figure}[htb]
  \centering
  \centerline{\includegraphics[width=\linewidth]{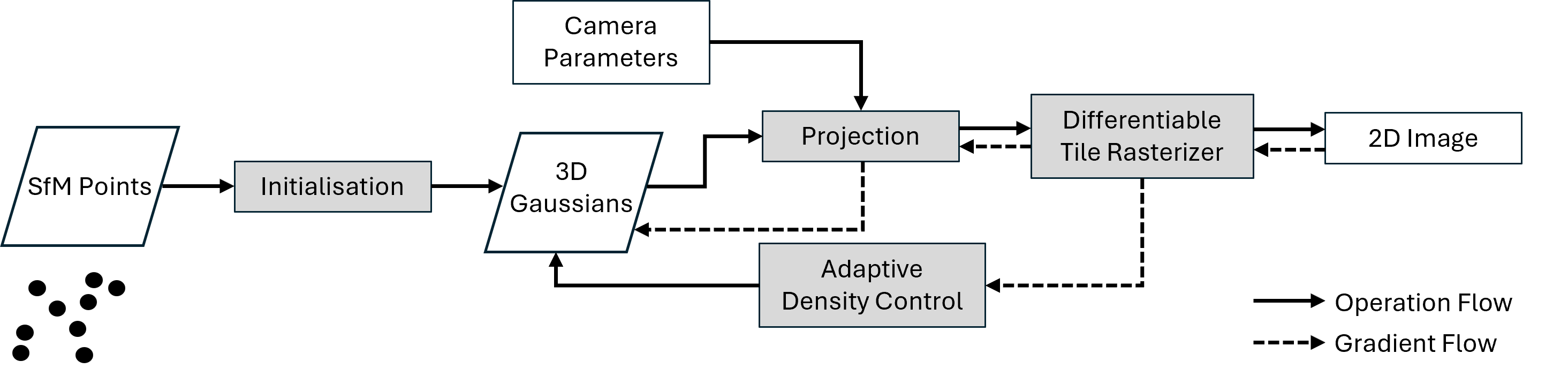}}
  \caption{Typical learning/optimisation process for producing a 3DGS model from an initial set of multi-view images, after the SfM points have been obtained from these images.}
  \label{fig:learning_pipeline}
\end{figure}

The typical way to generate a 3DGS model is to first capture a sufficiently large number of 2D images of the 3D object or scene of interest from a variety of viewpoints in 3D space, and then to pass this collection of images to an automated system that learns a 3DGS model of the scene by using a Stochastic Gradient Descent optimisation~\cite{Kerbl2023}. To initialise the learning process, a sparse 3D point cloud of the captured scene is usually estimated via a Structure-from-Motion (SfM) technique~\cite{Schonberger2016} applied on the input images, and these initial points then go through the optimisation process described in~\cite{Kerbl2023} to produce the final set of 3D Gaussian splats, as shown in Fig.~\ref{fig:learning_pipeline}.

\begin{figure}[htb]
  \centering
  \centerline{\includegraphics[width=\linewidth]{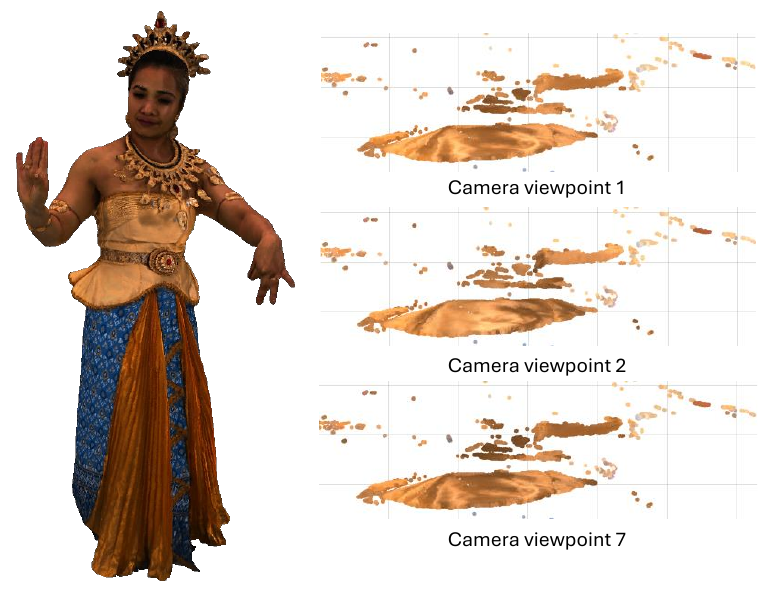}}
  \caption{One frame of the plenoptic \textit{Thaidancer} point cloud~\cite{Krivokuca2018} and part of her blouse as seen from different captured viewpoints. The full 300-frame video of this dataset can be found in the supplementary materials attached with this paper.}
  \label{fig:Thaidancer_mv_colours}
\end{figure}

In a \textit{plenoptic point cloud} dataset, such as the 8iVSLF~\cite{Krivokuca2018}, for each 3D point, or voxel (\textit{vo}lumetric pi\textit{xel}), there exists not one (R, G, B) colour triplet per point, but a \textit{set} of (R, G, B) vectors, each of which is associated with a different viewing direction in 3D space. That is, for each point $p_i$ in a set of points $\{ p_i \mid i \in [1, N_p] \}$ in $\mathbb{R}^3$, with spatial coordinates $(x_i,y_i,z_i)$, its geometry-colour vector representation is: 
\begin{equation}
p_i = \big[ x_i,\, y_i,\, z_i,\, R_i^{1},\, G_i^{1},\, B_i^{1},\, \ldots,\,
R_i^{N_c},\, G_i^{N_c},\, B_i^{N_c} \big],
\end{equation}
where $N_c$ is the number of camera viewpoints used to capture the 3D object or scene, and $\big[R_i^{1},\, ...,\, B_i^{N_c} \big]$ is the point's \textit{plenoptic}, or \textit{multi-view}, colour vector. Fig.~\ref{fig:Thaidancer_mv_colours} illustrates one frame of the dynamic point cloud sequence, \textit{Thaidancer}, from~\cite{Krivokuca2018} (consisting of 300 frames), together with examples of the different colour values on the dancer’s gold blouse as seen from some of the different captured viewpoints. We see that these colours change depending on the viewing position, due to the reflective nature of the blouse material. This clearly illustrates the usefulness of having \textit{view-dependent} colours per point, for realistically representing light reflections off specular (non-Lambertian) surfaces. 

\section{Proposed Transcoding Pipeline}
\label{sec:details_proposal}

The ideas described in this paper may essentially be considered methods of \textit{transcoding} a 3D plenoptic point cloud or mesh into a 3DGS model. The 3DGS model may then be passed to a dedicated codec framework (outside the scope of this paper), independently of how it was generated. The different steps of our transcoding pipeline are detailed in the following sub-sections, in order. Our goal here is to demonstrate that our methodology can be implemented by using existing and readily available software tools, such that an interested practitioner may easily replicate our results or generate their own transcoded models. In this way, we hope that our work will provide a practical starting point for more research in this direction.

\subsection{Generating multi-view images}
\label{ssec:mv_images}

\begin{figure}[htb]
  \centering
  \centerline{\includegraphics[width=0.8\linewidth]{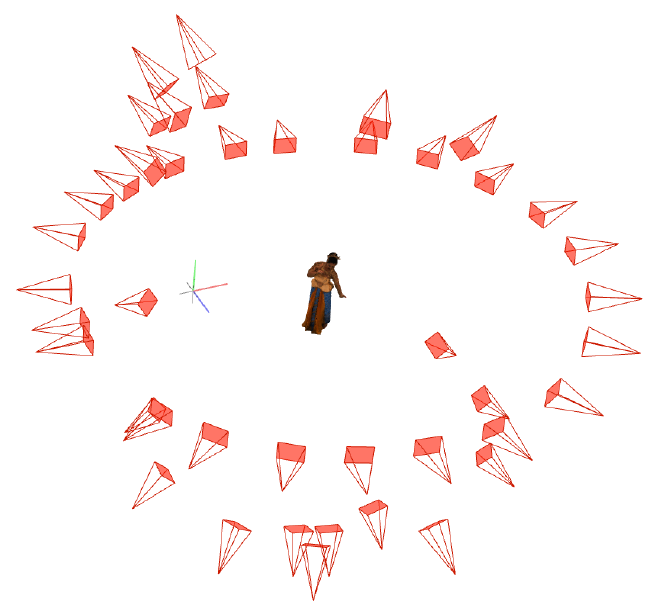}}
  \caption{Example distribution of camera viewpoints chosen to render multi-view images of an object-centric dataset.}
  \label{fig:Thaidancer_example_viewpoints}
\end{figure}

\begin{figure}[htb]
  \centering
  \centerline{\includegraphics[width=\linewidth]{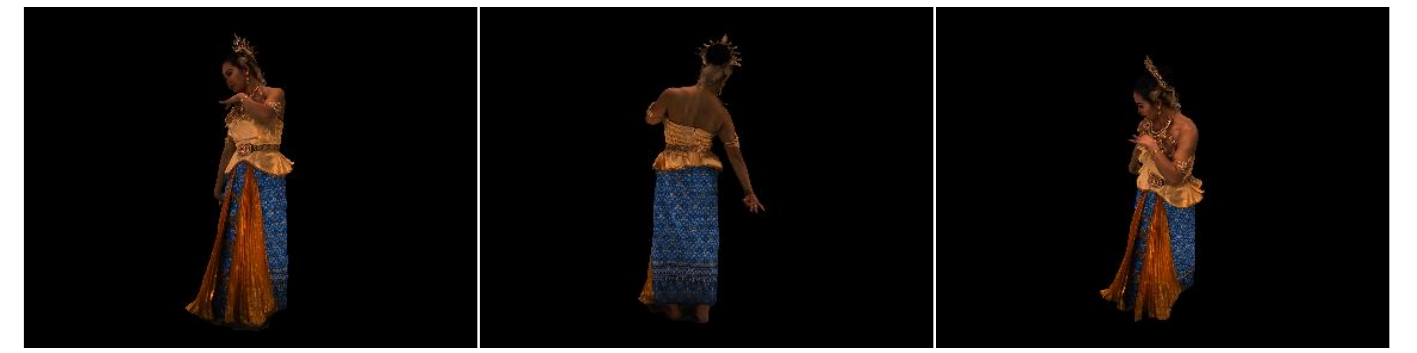}}
  \caption{Example multi-view images (4080x3060 pixels) generated for \textit{Thaidancer}~\cite{Krivokuca2018}, by using the MPEG PccAppRenderer~\cite{Ricard2018}.}
  \label{fig:Thaidancer_images}
\end{figure}

The first step in our proposed methodology is to generate multi-view images of the available point cloud or mesh, since the original captured images are unavailable. This can be done by using a suitable renderer, such as the MPEG PccAppRenderer~\cite{Ricard2018} (or other), which is able to render plenoptic point clouds with varying colours per viewpoint. We then need to choose a sampling of $N_c$ camera viewpoints that sufficiently covers the object or scene in question. In general, the exact number $N_c$ and sampling density of the viewpoints will require some experimentation for any given input model, but for object-centric datasets such as the 8iVSLF, an approximately spherical arrangement of cameras pointing inwards and approximately equally distributed, as shown in the example in Fig.~\ref{fig:Thaidancer_example_viewpoints}, tends to work well, since this is how the subjects are usually captured in the first place~\cite{Krivokuca2018}. To display the correctly varying colours per viewpoint, the PccAppRenderer uses an interpolation method, which depends on the position of the point, camera, and viewer, as follows. For each point $j$ in the input point cloud, and for each of the chosen camera viewpoints $i$, the renderer computes a dot product:
\begin{equation}
w_i = (\vec{C}_i - \vec{P}_j) \cdot (\vec{V} - \vec{P}_j),
\end{equation}
where $\vec{C}_i$ is the position of camera $i$, $\vec{P}_j$ is the position of point $j$, and $\vec{V}$ is the position of the viewer. This dot product measures the alignment between the viewing direction and the camera direction relative to the point $j$. To ensure only positively aligned contributions, the weight is scaled as:
\begin{equation}
w_i = \max(0, w_i^{\,n}),
\end{equation}
where $n$ is a sharpness parameter that controls the influence of each camera ($n=10$ by default). The colour of point $j$ as seen from viewpoint $i$ is then computed as a weighted average:
\begin{equation}
\mathrm{Colour}_j = 
\frac{\sum_i w_i \cdot \mathrm{Colour}_i}{\sum_i w_i}.
\label{eq:pccapprend_interp}
\end{equation}
The interpolation in~\eqref{eq:pccapprend_interp} ensures that the cameras that are more aligned with the viewer's viewing direction contribute more significantly to the final colour. We save the rendered 2D image from each of the chosen viewpoints, with the colours as seen from that viewpoint. Fig.~\ref{fig:Thaidancer_images} shows some example images generated for \textit{Thaidancer}, by using the PccAppRenderer and the camera arrangement shown in Fig.~\ref{fig:Thaidancer_example_viewpoints}. It is important to use a high enough image resolution to obtain good quality images that will be usable for the subsequent 3DGS model generation, but this requires some experimentation as the appropriate resolution will depend on the quality and density of the input 3D point cloud or mesh.

\subsection{Obtaining extrinsic and intrinsic camera parameters}
\label{ssec:cam_params}

For each of the $N_c$ camera viewpoints chosen for rendering, we also need to know the camera \textit{extrinsic} and \textit{intrinsic} parameters, in order to obtain the correct transformations between the “world” (captured scene) coordinate system and the camera coordinate system that is used for projection onto a 2D (pixel) plane to obtain the rendered images. The extrinsic parameters define the location and orientation of a camera in the 3D capture space and are collectively known as the camera’s \textit{pose}. The intrinsic parameters are specific to a camera design and define the camera’s geometric properties; they determine how the internal 3D-to-2D projection is done between the 3D camera coordinates and the pixel coordinates in the rendered image. The \textit{extrinsic} parameters for each camera are:
\begin{itemize}[nosep]
    \item Its \textit{position}, specified by a vector $(\vec{x}_w,\, \vec{y}_w,\, \vec{z}_w)$ in the 3D coordinate system of the scene (“world”),
    \item Its \textit{orientation}, i.e., a 3D rotation around the axes of the world coordinate system, either specified by 3 Euler angles or with a 3D unit \textit{quaternion} representation.
\end{itemize}

\begin{figure}[htb]
  \centering
  \centerline{\includegraphics[width=\linewidth]{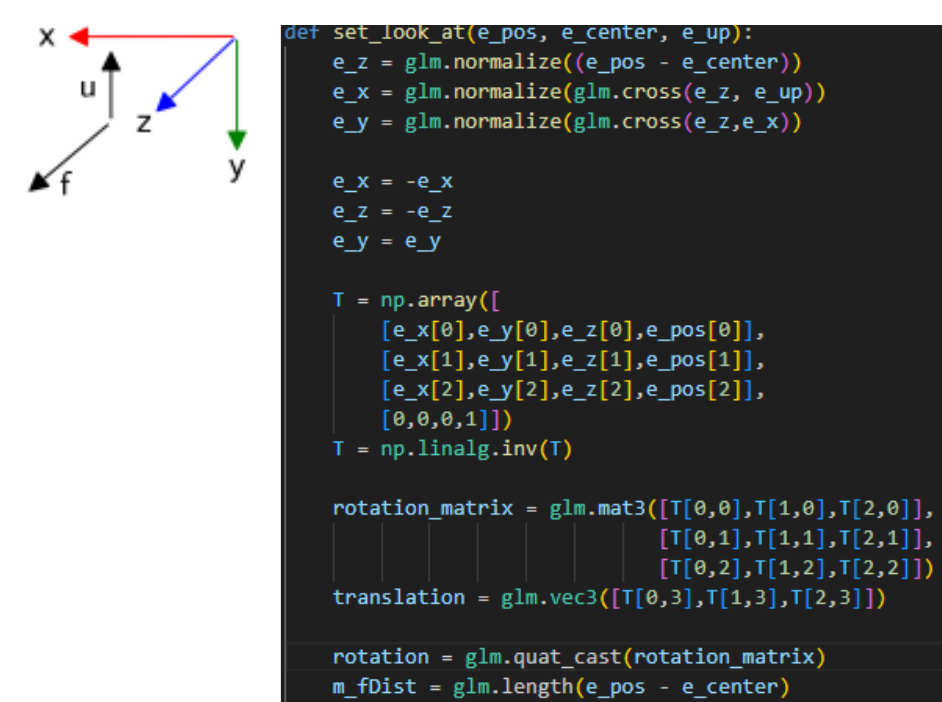}}
  \caption{COLMAP's~\cite{Schonberger2016} coordinate system (left) and our Python code (right) for transforming the PccAppRenderer's~\cite{Ricard2018} coordinate system to COLMAP's.}
  \label{fig:coord_system_transform}
\end{figure}

\noindent The origin of the camera’s coordinate system is at its optical centre, $[c_x,c_y]$, and its x- and y-axes define the image plane. Our ideas in this paper are independent of the method by which the camera extrinsic parameters are obtained, as this depends on the chosen rendering and 3DGS learning pipelines. But it is important that the transform between the coordinate system used by the renderer to generate the multi-view images and the coordinate system used for the 3DGS model learning is known, as we need the input 3D model to be in the same coordinate system in both. For all the results provided in this paper, the extrinsic parameters have been obtained by transforming the viewport coordinates in the PccAppRenderer to COLMAP's~\cite{Schonberger2016} “camera” coordinate system. Fig.~\ref{fig:coord_system_transform} shows the COLMAP coordinate system ($u$ and $f$ represent the \textit{up} and \textit{front} vectors of a camera, respectively) and our Python code for the required transformation. The \textit{intrinsic} parameters for each of the chosen $N_c$ cameras are:
\begin{itemize}[nosep]
    \item The focal length, $(f_x,f_y)$, in pixels,
    \item The optical centre (principal point), $[c_x,c_y]$, in pixels,
    \item The skew coefficient, $s$, which is non-zero if the image axes are not perpendicular.
\end{itemize}
Sometimes other parameters, e.g., distortion coefficients, are also taken into account, for correcting lens distortions in non-pinhole camera models. Our ideas in this paper are independent of the exact camera model used and the method by which the intrinsic parameters are obtained or estimated.

\subsection{Initialising the 3DGS learning process}
\label{ssec:initialisation}

Once the multi-view images have been generated and the associated camera parameters have been obtained, as explained above, the next step is to initialise the 3DGS learning process. A good initialisation is important, because it can heavily influence the quality of the final 3DGS model and the number of learning/optimisation iterations required to obtain this model. We propose two methods for this initialisation, starting from our generated multi-view images:
\begin{enumerate}[label=\roman*., nosep]
    \item Passing the images to a fully automated system for 3DGS learning (e.g., \textit{gsplat}~\cite{Ye2024}, or other) and using the default initialisation process from SfM, as shown in Fig.~\ref{fig:learning_pipeline}, or
    \item Using a custom initialisation that takes into account the geometry of the already available 3D point cloud or mesh. 
\end{enumerate}  
\noindent Both initialisation possibilities are described below.

\subsubsection{Using the default initialisation}
\label{sssec:init_default}

In the default initialisation, we rely on the automated SfM process (e.g., as used in~\cite{Kerbl2023}) to estimate the initial points from which to start the 3DGS learning. While this is convenient, as it requires no additional effort from the user, it usually produces worse results than when using a more intelligent, custom initialisation, and can be problematic when we wish for the generated 3DGS model not to stray (far) from the initial point cloud or mesh surface. 

\subsubsection{Using a custom initialisation}
\label{sssec:init_custom}

\begin{figure}[htb]
  \centering
  \centerline{\includegraphics[width=0.9\linewidth]{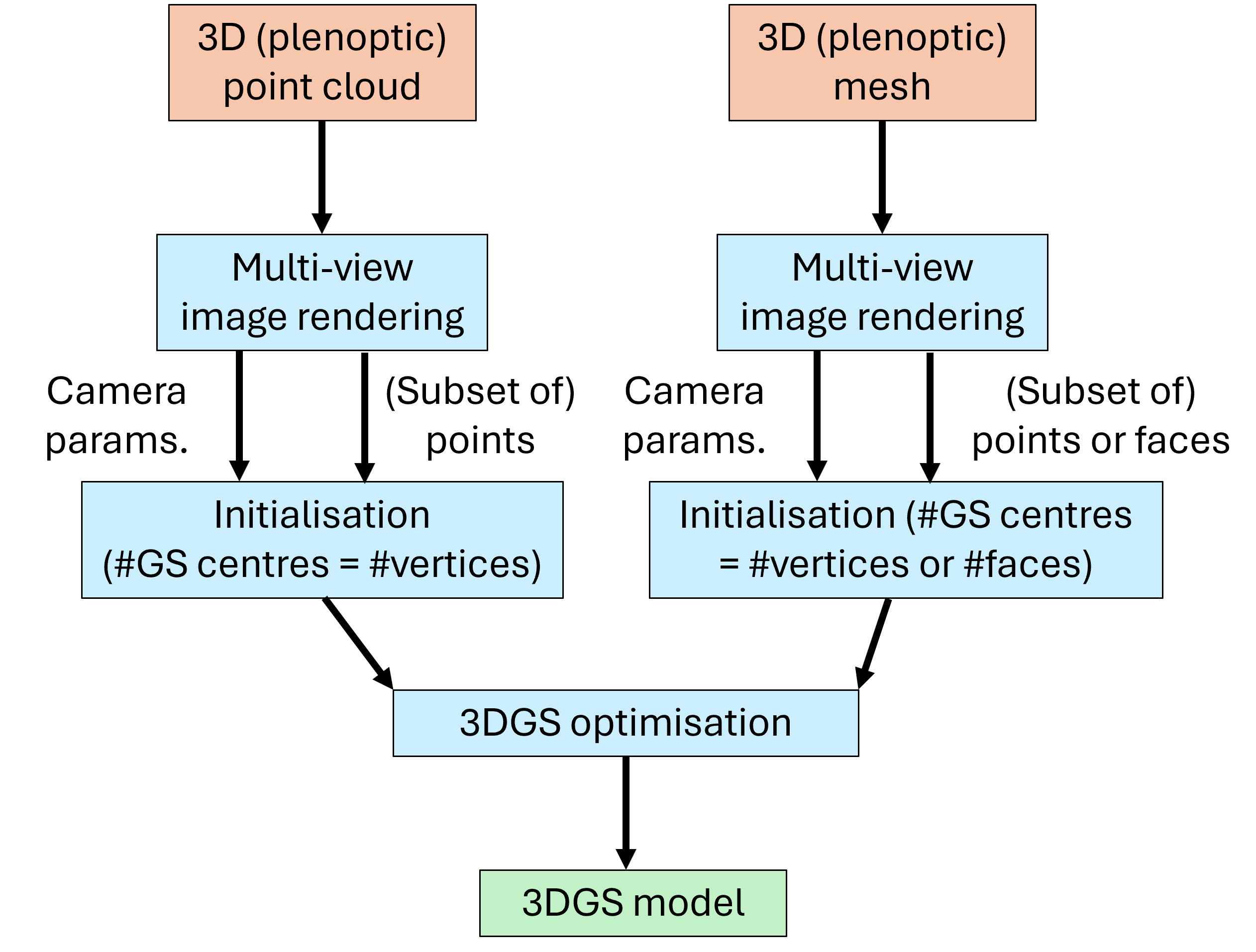}}
  \caption{The proposed methodology when using our custom initialisation for 3DGS model learning/optimisation.}
  \label{fig:custom_init_process}
\end{figure}

Fig.~\ref{fig:custom_init_process} summarises our proposed methodology when using our custom initialisation. As shown in Fig.~\ref{fig:custom_init_process}, whether the original input model is a 3D point cloud or mesh, the same multi-view image generation (e.g., as described in Sec.~\ref{ssec:mv_images}) is applied. Then, for the initialisation process to begin the 3DGS learning, we pass in the camera parameters (e.g., as described in Sec.~\ref{ssec:cam_params}) for the rendered camera viewpoints, as well as (a subset of) the points ($(x,y,z)$ vertices) that are in the input 3D model. If the input is a 3D mesh, we may alternatively use the centre point of each mesh face (or the centres of a chosen subset of faces), or some other known points on the object’s surface. The use of some known points on the input model surface is a key component of our custom initialisation, because this ensures that the generated splats do not move far (or at all) from the original 3D object surface. This can be important in applications that require retaining the original 3D object geometry, since it is known that 3DGS models on their own (i.e., those generated without any constraints related to the captured scene geometry) do not usually provide a natural (clean) surface structure, as they are learned according only to error metrics measured on the rendered 2D images of the 3D object or scene, without any 3D geometric error metrics.

Normally, in the “adaptive density control” part of the 3DGS learning pipeline (Fig.~\ref{fig:learning_pipeline}), splats that are too small to cover a designated area are cloned, while splats that are too large are split into smaller splats. Densification is used to add points to the 3DGS model, because the SfM process produces a fairly sparse point cloud to begin with. But if the input point cloud is already a dense and ``clean" enough representation of the underlying 3D object surface, as is the case for the 8iVSLF dataset, then we can deactivate the adaptive density control, as we already have the correct point density and placement for the final splats. This also has the benefit of reducing the learning time, which may be significant in large scenes where many splats need to be processed. In our custom initialisation, to ensure that the initial points correspond exactly to the centres of the splats in the final 3DGS model, we can set to 0 the “position learning rate”~\cite{Kerbl2023} for the splat centres, while retaining a learning rate $> 0$ for the other splat attributes. Examples of the results of such a process are shown in Sec.~\ref{sec:results}. Note that if the input 3D model is of a poorer quality (e.g., is noisy, sparse, or has holes), then we should not deactivate the adaptive density control, so as to ensure a better quality 3DGS model at the end.

\section{Results}
\label{sec:results}

Here we present the 3DGS models obtained by applying our proposed transcoding pipeline to the 8iVSLF point cloud sequence, \textit{Thaidancer}~\cite{Krivokuca2018}. Similar results can also be obtained on the other 8iVSLF models, but we do not show these here due to space limitations. \textit{Thaidancer} is the most interesting 8iVSLF model for transcoding to 3DGS, as it contains the most obviously varying colours and specular details as seen from different viewpoints. It is also the only \textit{dynamic} point cloud in 8iVSLF, so the results shown here are actually samples of our tests on all 300 frames; our generated 3DGS models for the other frames can be seen in the video that is included as supplementary material with this paper.

\begin{figure}[h!]
  \centering
  \centerline{\includegraphics[width=\linewidth]{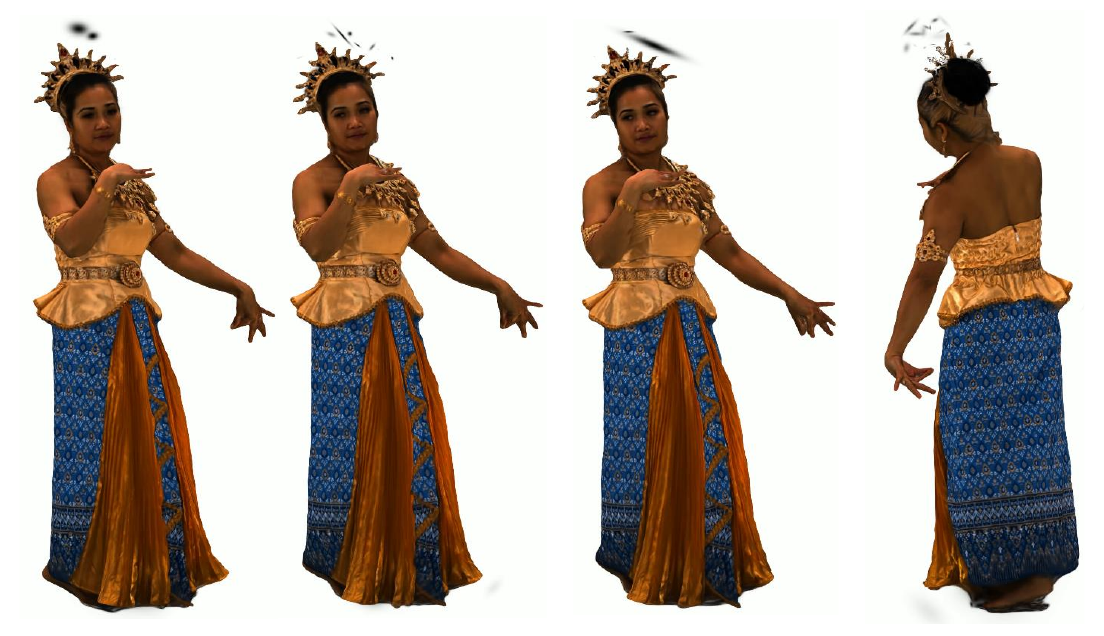}}
  \caption{3DGS models (containing $\sim115\text{k}$ splats per frame) produced with our proposed transcoder, when using the default SfM-based initialisation and 15k learning iterations.}
  \label{fig:Thaidancer_COLMAP_15k}
\end{figure}

Fig.~\ref{fig:Thaidancer_COLMAP_15k} shows the generated 3DGS models for a few frames of \textit{Thaidancer}, obtained when using our transcoding pipeline and the \textit{default} learning initialisation (Sec.~\ref{sssec:init_default}). This corresponds to using the COLMAP~\cite{Schonberger2016} SfM method, as part of the gsplat~\cite{Ye2024} system, to estimate a sparse 3D point cloud for initialising the 3DGS learning, after passing in our images generated by the MPEG PccAppRenderer~\cite{Ricard2018} from the 44 camera viewpoints shown in Fig.~\ref{fig:Thaidancer_example_viewpoints}. We see that the 3DGS models are of very good visual quality, but even after 15,000 learning iterations (and still after 30,000, not shown here), there are visible artifacts in the form of black “floater” splats above the dancer’s head and near her feet. This happens because some of the generated splats do not align well with the object’s surface, as the SfM points are not constrained to align with the original point cloud geometry. These floaters are the same colour as the image background, but there is actually no background in the original point clouds -- this background is produced by the renderer when creating the multi-view images (see Fig.~\ref{fig:Thaidancer_images}). While only certain frames are shown as examples in Fig.~\ref{fig:Thaidancer_COLMAP_15k}, the black floater splats are actually present in all 300 frames of the generated \textit{Thaidancer} 3DGS sequence. This can be seen in the video that is provided as supplementary material with this paper.

\begin{figure}[h!]
    \centering
\centerline{\includegraphics[width=0.7\linewidth]{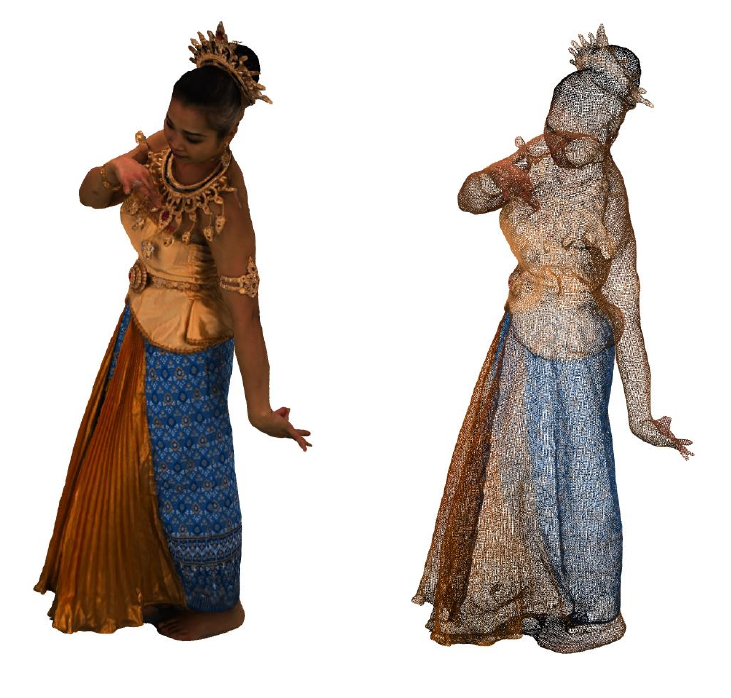}}
      \caption{(Left) One frame of the original \textit{Thaidancer}~\cite{Krivokuca2018} point cloud sequence ($\sim3\text{M}$ voxels); (Right) Sparse mesh ($\sim250\text{k}$ vertices).}
    \label{fig:Thaidancer_mesh_250kPts}
\end{figure}

\begin{figure}[h!]
  \centering
  \centerline{\includegraphics[width=\linewidth]{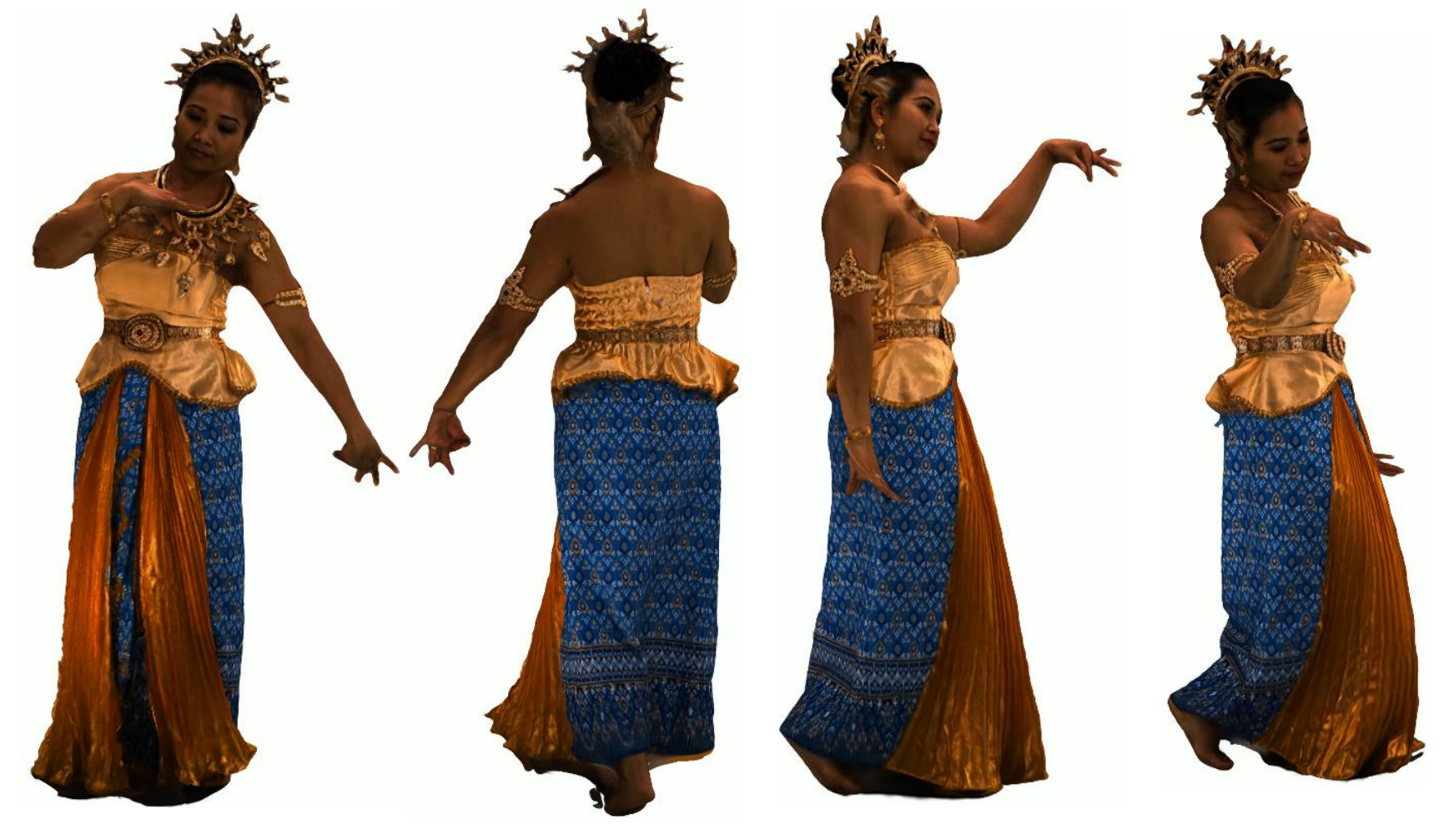}}
  \caption{3DGS models (containing $\sim250\text{k}$ splats per frame) produced with our proposed transcoder, when using our custom initialisation and 5k learning iterations.}
  \label{fig:Thaidancer_ours_5k}
\end{figure}

In contrast, such floater artifacts (as seen in Fig.~\ref{fig:Thaidancer_COLMAP_15k}) do not appear when using our \textit{custom} initialisation (Sec.~\ref{sssec:init_custom}), as shown in Fig.~\ref{fig:Thaidancer_ours_5k}, because in this case the splats are guaranteed to align with points on the original 3D model surface. For example, to generate the 3DGS models in Fig.~\ref{fig:Thaidancer_ours_5k}, we started with much sparser, \textit{meshed} versions of the original dense point clouds, as shown in Fig.~\ref{fig:Thaidancer_mesh_250kPts}. We used the vertices of these meshes as the initial points from which to start the 3DGS learning, so the number of generated splats is equal to the number of mesh vertices in the corresponding frame of the sequence, and the splats are centred on those vertices. We see in Fig.~\ref{fig:Thaidancer_ours_5k} that the visual quality of the generated 3DGS models is again very good, basically visually indistinguishable from the original, much denser point cloud (e.g., compare to Fig.~\ref{fig:Thaidancer_mesh_250kPts} (Left)), while also converging several times faster when using our custom initialisation (5k iterations for the examples in Fig.~\ref{fig:Thaidancer_ours_5k}) compared to when using the default SfM-based initialisation (15k iterations for the examples in Fig.~\ref{fig:Thaidancer_COLMAP_15k}). Subjectively, the comparison between our results in Figs.~\ref{fig:Thaidancer_COLMAP_15k} and~\ref{fig:Thaidancer_ours_5k} demonstrates that, for a comparable visual quality, our custom initialisation leads to cleaner, more surface-bound 3DGS models than the SfM initialisation. Note that we intentionally show different viewpoints between Figs.~\ref{fig:Thaidancer_COLMAP_15k} and~\ref{fig:Thaidancer_ours_5k}, in order to demonstrate that the generated 3DGS outputs maintain a consistently high visual quality across a range of views, rather than spotlighting only a few favourable viewpoints. The main observations that we wish to highlight are the high visual quality of the produced models, regardless of which initialisation or viewpoint is used, and the lack of floater artifacts when using our custom initialisation. Also note that the different numbers of splats between Figs.~\ref{fig:Thaidancer_COLMAP_15k} and~\ref{fig:Thaidancer_ours_5k} is not important here, as these numbers are only examples that can be adjusted depending on the user's 3DGS model requirements. The floaters visible in Fig.~\ref{fig:Thaidancer_COLMAP_15k} are not related to the smaller number of splats. To facilitate a more complete comparison for the reader, the supplementary material attached with this paper includes full videos of the generated 3DGS sequences corresponding to Figs.~\ref{fig:Thaidancer_COLMAP_15k} and~\ref{fig:Thaidancer_ours_5k}, as well as the original point cloud sequence from~\cite{Krivokuca2018}, viewed from different viewpoints.

The reader will notice that we do not provide objective error metric results in this paper. This is because reliable objective evaluation is not always possible in our setting. In particular, the SfM-based initialisation produces floater splats that are the same colour as the background of the generated multi-view images (see Fig.~\ref{fig:Thaidancer_COLMAP_15k}), so these floaters are not penalized by standard metrics (since they do not register as artifacts), leading to misleading metric results. Addressing this would require modifying the evaluation pipeline or designing new metrics, which is beyond the scope of this paper. Pruning the floaters is a possibility, but this would not constitute a fair comparison to our custom initialisation, which does not produce these floaters in the first place. More importantly, our goal in this paper is not comparative benchmarking (since, as noted in Sec.~\ref{sec:intro}, other comparable transcoding methods currently do not exist), but to provide a first proof of concept of a practical pipeline for generating 3DGS models from plenoptic data. As in many 3DGS applications, visual quality for human observers is our primary target, and existing objective metrics are known to correlate poorly with perceived quality. The absence of such metrics is therefore intentional here.

The fact that, with our proposed transcoder, we are able to produce 3DGS models with many fewer splats than points in the corresponding original point clouds, for comparable visual quality, regardless of which of the two initialisation methods is used, demonstrates another interesting benefit of transcoding existing plenoptic point cloud datasets into 3DGS models. The flexible sizes and shapes of Gaussian splats can allow us to represent the same 3D object surface with many fewer points (splats) than when using a dense point cloud where every point must have the same fixed size to achieve a similar perceived visual quality. So, in applications where the precise 3D geometry is not important, but the goal is to have a perceptually good-quality rendered output, our transcoding may serve as a useful pre-processing step to a compression algorithm, as it significantly reduces the number of points that need to be encoded. For example, for the first frame of the original \textit{Thaidancer} point cloud sequence~\cite{Krivokuca2018}, the raw (uncompressed) binary PLY (\textit{Polygon File Format}) file has a size of around 201MB, whereas the raw PLY for the first frame of our 3DGS sequence corresponding to the results in Fig.~\ref{fig:Thaidancer_ours_5k} has a size of around 61MB. Therefore, the transcoding step already provides a preliminary compression for a comparable visual quality.

\section{Conclusion}
\label{sec:conclusion}

In this paper, we proposed the first end-to-end pipeline for transcoding 3DGS models from existing plenoptic 3D point cloud (or mesh) models, when the original multi-view images of the captured 3D data are not available. Results demonstrate that our proposed system is able to produce high-quality 3DGS models, with many fewer splats than the number of points in an original dense point cloud. Our proposed custom initialisation for the 3DGS model learning also ensures that the final 3DGS model remains closely aligned with the original 3D point cloud or mesh surface -- this leads to better visual quality and faster model convergence than when using the typical SfM-based learning initialisation. We show how the proposed methodology can be implemented using existing software tools, to enable interested practitioners to easily reproduce our results.

\bibliographystyle{IEEEbib}
\bibliography{strings,refs}

\end{document}